\documentclass[preprint2]{aastex}
\usepackage{epsf}

\tighten

\newcommand{\gtsim}{\ {\raise-0.5ex\hbox{$\buildrel>\over\sim$}}\
}
\newcommand{\ltsim}{\ {\raise-0.5ex\hbox{$\buildrel<\over\sim$}}\
}
\def \etal{{\it et~al.~}}

\def\simlt{\lower.5ex\hbox{$\; \buildrel < \over \sim \;$}}
\def\simgt{\lower.5ex\hbox{$\; \buildrel > \over \sim \;$}}

\shortauthors{NEWMAN AND DAVIS}
\shorttitle{Measuring the Cosmic Equation of State}

\begin{document}

\title{{Measuring the Cosmic Equation of State with Counts of Galaxies  II: Error Budget for the DEEP2 Redshift Survey}}

\author{Jeffrey A. Newman and Marc Davis\altaffilmark{1}} 

\affil{Department of Astronomy, University of California, Berkeley, CA 94720-3411}
\email{jnewman@astro.berkeley.edu, marc@astro.berkeley.edu}

\altaffiltext{1}{Also Department of Physics, U.C. Berkeley}
\vskip -12pt
\begin{abstract}
In a previous paper, we described a new variant on the classical
``d$N$/d$z$'' test which could be performed using data from the next
generation of redshift surveys.  By studying the apparent abundance of
galaxies as a function of their circular velocity or velocity
dispersion, rather than luminosity, it is possible to avoid many of
the uncertainties of galaxy evolution while using quantities that may
be measured directly.  In that work, we assumed that counting
statistics would dominate the resulting errors.  Here, we present the
results of including cosmic variance and determine the impact of
systematic effects on attempts to perform the test with the upcoming
DEEP2 Redshift Survey.  For the DEEP2 survey geometry, cosmic variance
yields errors roughly twice those predicted from Poisson statistics.
Through Monte Carlo simulations we find that if the functional form,
but not the strength, of any of the major systematic effects (baryonic
infall, velocity errors, and incompleteness) is known, the free
parameter may be determined from the observed velocity function.  The
systematic may then be corrected for, leaving a much smaller residual
error.  The total uncertainty from systematics is comparable to that
from cosmic variance, but correlated amongst redshift bins.  Based on
these analyses, we present error budgets for a d$N$/d$z$ measurement
with DEEP2 and determine the resulting constraints on cosmological
parameters.  We find that the uncertainty in the cosmic equation of state
parameter $w$ are $\sim2\times$ higher than previously derived,
providing a measurement much stronger than any available today
but weaker than some other proposed tests.

\end{abstract}

\keywords{cosmological parameters, cosmology: observations, galaxies: high-redshift,galaxies: fundamental parameters}

\section{Introduction}

In a previous paper (Newman \& Davis 2000, hereafter ND00), we
described a new variant on the classical ``d$N$/d$z$'' test which
could be performed using data from the next generation of redshift
surveys.  This classical technique determines the evolution of the
cosmic volume element by measuring the redshift distribution of a
tracer whose number density is known, providing
constraints on fundamental cosmological parameters.  In the past, the
abundance of galaxies was used to perform this test under the
assumption that the total comoving number density of galaxies
integrated over all luminosities is independent of redshift (with further
assumptions about the luminosity function; e.g. Loh \& Spillar 1986).
Such assumptions may be suspect, but improving upon them would
require a reliable, comprehensive theory of galaxy formation and
evolution.

Instead, ND00 suggests that by measuring the apparent abundance of
galaxies as a function of their linewidth or velocity dispersion
(normalized to the abundance at $z\sim 0$), it is possible to obtain a
more reliable measurement of the volume element, exploiting the
simplicity of the dark matter halo velocity function.  Current and
upcoming large redshift surveys with moderately high spectroscopic
resolution will make this possible.  The DEIMOS/DEEP Redshift Survey
(hereafter, ``DEEP2'') is particularly well suited for the
technique. This project is a major effort to observe distant galaxies
using the new DEIMOS spectrograph on the Keck telescope, scheduled to
be installed in early 2002.  Its goal is to obtain high quality
spectra of 50,000 galaxies selected to have minimum redshift $z>0.7$
(the ``1HS'', or 1-hour survey, so named because of the expected
exposure time per slitmask) and spectra of 5,000 such galaxies in
selected regions to a fainter limiting magnitude (the ``3HS'', or
3-hour survey).  DEEP2 will obtain data characterizing galaxies
and large-scale structure at $z\sim1$ comparable in quality to
what is currently available for $z=0$.  Although we have much information
on the local universe, we know little about how it has reached its
present state; this survey has been designed to address this major gap
in our knowledge.

In our previous work, we assumed that the uncertainties in a d$N$/d$z$
measurement using DEEP2 would be dominated by Poisson statistics and
found that it could provide strong constraints on the cosmic equation
of state parameter of quintessence-like models, $w= P /\rho$.  Because
the velocity function of dark matter halos is so simple (a nearly
perfect power-law at galactic scales), we presumed that any systematic
effects would leave a clear signature and could be removed or avoided,
and that counting statistics would therefore dominate the
uncertainties.  However, we did not consider the fundamental limit to
the accuracy of any count imposed by cosmic variance.  In this paper,
we calculate its impact on the DEEP2 survey.

Since ND00, a number of papers suggesting that systematic
effects would compromise our proposed test have appeared in the
literature (e.g. Bullock et al. 2000), though again without actually
evaluating their impact on realistic methods of measuring d$N$/d$z$.
This paper addresses that gap in our previous work; we
have performed quantitative tests of the impact systematic effects
should have on our ability to measure d$N$/d$z$ with DEEP2.
However, we must do so without the information that will be available
at the completion of the survey.  For instance,
semi-analytic and N-body estimates of the impact of baryonic infall on
the potential well depths of galaxies are still developing
(e.g. Kochanek \& White 2001), and comparisons between HI rotation curves
and O[II] linewidths for DEEP2 galaxies may be possible in the
future, but are unavailable to us now.

Rather than attempting to foresee all that the next few years will
bring, we have instead chosen to evaluate the impact of major
systematic effects by applying ``toy'' models of their impact to a
semi-analytic prediction of the dark-matter halo velocity function (a
la ND00).  The semi-analytic methods used duplicate the general
behavior seen in N-body simulations.  The toy models used each have
one free parameter that is treated as unknown going into the analysis.
We then generate Monte Carlo realizations of the velocity function
affected by these model systematics and determine the resulting errors
in measuring both the free parameter and the abundance of dark halos
in such datasets.  Thus, for any value of the free parameter we can
determine the excess variance in a d$N$/d$z$ measurement due to the
residual uncertainty from that systematic effect after removal.  So
long as the toy models duplicate the major features of the real
systematics, we should obtain reasonable estimates of the actual
errors from DEEP2.  In $\S$ 2 of this paper, we calculate the impact
of cosmic variance; in $\S$ 3, we determine the residual errors that will
result from systematic effects; and in $\S$ 4, we
present the constraints on cosmological parameters that a measurement
of d$N$/d$z$ from the velocity function of DEEP2 galaxies would
provide.

\section{Cosmic Variance}

By definition, the spatial correlations of galaxies imply that they
are not distributed independently and thus cannot obey simple
Poisson statistics.  Instead, in a given volume there will be an
excess variance
\begin{equation}
\sigma_{CV}^2={1 \over V^2} \int_V \xi(x_1,x_2) d^3 x_1 d^3 x_2,
\end{equation}
where  $\xi$ is the
two-point correlation function of galaxies and $V$ is the volume in which we wish
to calculate the variance.  This integral may more easily be
evaluated in Fourier space, where we obtain
\begin{equation}
\sigma_{CV}^2={1 \over {8 \pi^3}} \int P(k) |\tilde{W}(k)|^2 d^3 k,
\end{equation}
where $\tilde{W}(k)$ is the Fourier transform of a spatial window
function that is 1 inside the volume and zero elsewhere and $P(k)$ is
the power spectrum of the galaxy distribution.  To evaluate the
integral, we use the CDM-like power spectrum of Bardeen et al. (1986)
with $\Gamma=0.25$ and slope $n=1$, consistent with recent
observations.

We are interested specifically in the effect of cosmic variance upon
measurements of d$N$/d$z$ with DEEP2 (as the volume observed in the
next generation of $z\sim 0$ surveys is much greater, cosmic variance
in local comparison samples should be comparatively negligible).
Because the CFH12k camera used in imaging has a
30$\arcmin\times40\arcmin$ field of view, the most efficient geometry
for each of the four fields to be surveyed \cite{deepwfsc} is one
which is $30\arcmin$ in the shortest direction.  Based on observing
time constraints, we further expect each field to cover $120\arcmin$
in the long direction on the sky; the required imaging is nearly
complete.  If we assume an $\Omega_m$=0.3 $\Lambda$CDM model, this
corresponds to $20.2 \times 80.8 h^{-1}$ Mpc comoving at $z=1$, the
expected median redshift for the survey.  We have calculated the
cosmic variance for each field assuming that the volume surveyed is a
rectangular solid extending 1300 $h^{-1}$ Mpc comoving in the redshift
direction (corresponding to $0.7<z<1.5$).

For a rough estimate, we could assume that the power spectrum of
galaxies has the same normalization as clusters today, applying the
$\sigma_8$ measurement of Borgani et al. (1999), which yields 0.96 for
an $\Omega_m=0.3$ LCDM model (for quintessence models, we use the
relations of Wang and Steinhardt (1998) with $\Theta=0.075$).  We then
find that the uncertainty in a density measurement will be 5.9\% in
each of the four fields, or 2.9\% for the entire survey (as the fields
are widely spaced on the sky, there should be minimal correlation
between them at $z=1$, so the combined variance goes as
1/$N_{fields}$).

However, to calculate the variance in counts of galaxies, we must use
the power spectrum of galaxies at $z=1$ rather than clusters at $z=0$;
we therefore must renormalize to match $\sigma_8$ for galaxies.  A
measurement of the correlation length $r_0$ may be directly
transformed into a measurement of $\sigma_8$ if the correlation index
$\gamma$ is known (Peacock 1999); however, the correlation properties
of galaxies at $z\sim 1$ remain poorly known.  Observations suggest
that $r_0 \sim 2.5 h^-1$ Mpc (comoving; see Hogg et al. 2000), similar
to the correlation length of the dark matter at that redshift, but
semi-analytic models predict $r_0 \sim 4 h^{-1}$ Mpc (comoving) for
luminous galaxies at that redshift \cite{benson01}.  For $\gamma=1.8$
and $r_0 = 2.5,3.25,4$, $\sigma_8 = 0.474, 0.600, 0.723$.  Thus, we
expect cosmic variance in the DEEP volume to be somewhat lower than a
naive calculation would yield: 1.4 -- 2.2 \% rather than 2.9\%.  Since
DEEP2 will obtain linewidths for $\sim 10^4$ galaxies with linewidths,
the uncertainty due to cosmic variance will be roughly twice the
Poisson value.  In Figs. 1 and 2, we show the effect of changing the
survey geometry or areal coverage on the cosmic variance in overall
DEEP2 results.  The amount of area surveyed on the sky is clearly of
greater importance than the shape of the surveyed region.

Cosmic variance need not impose a permanent limit to the precision of
a DEEP2 d$N$/d$z$ measurement; if imaging of a wider area is available
sufficient to produce an accurate photometric redshift distribution
(even if information on individual galaxies remains uncertain), it
could be possible to renormalize the total abundances of galaxies in
the DEEP2 fields and reweight the abundances of galaxies with observed
linewidths.  Other redshift surveys extending to $z\sim1$,
particularly the VLT/VIRMOS survey, could also provide corrections.
However, cosmic variance would almost certainly remain an
insurmountable obstacle to efforts to measure the evolution in the
equation of state parameter $w$ via DEEP2 d$N$/d$z$ alone, since the
variance is higher in any subvolume than for the survey as a whole.

\section{Correction for systematic effects}

Like most astronomical studies, the measurements proposed by ND00
could be subject to a number of systematic effects.  If these
systematics are well-understood and their magnitude is known, it is
straightforward to correct for them when measuring the halo velocity
function.  However, few, if any, of the systematic effects that will
affect this measurement can yet be predicted accurately.
For instance, we can determine $a~priori$ what effect a given
luminosity-linewidth relation would have upon incompleteness in the
velocity function, but without sufficient measurements at $z \sim 1$
(which will not be possible until the DEEP2 survey is conducted), we
cannot determine the resulting uncertainties with precision.

Thus, rather than attempt to guess the magnitude of systematic effects
that will be better understood five years from now and calculate the
errors in a measurement of d$N$/d$z$ that would result, we have
attempted to determine to what degree we can measure and correct for
systematics using the signatures they leave on the velocity function
at $z\sim1$.  In many cases, this is a pessimistic assumption; for
instance, we will be able to measure our errors in determining
linewidths by comparison to absorption line measurements of DEEP2
galaxies, rather than having to infer them from the velocity function
alone.  However, this technique has the significant advantage that we
do not need to understand perfectly either the intrinsic velocity
function of dark halos or any of the systematic effects to be able to
assess their impact, so long as we can define ``toy'' models for them
that replicate their major features and have the same free parameters
that must be determined from the observations.  For the intrinsic dark matter
velocity function, we use the semi-analytic method of ND00.  We have
tested models of three major systematic effects expected in our
measurements: the impact of baryonic infall on the circular velocities
of dark halos; the effects of incompleteness at low circular
velocities; and random errors in determining those circular velocities from observations.  We describe our models for these in more detail below.

As a further simplification, since all physical models of either the
halo velocity function or systematic effects simultaneously predict
their dependence upon redshift, we determine the impact of systematics
on d$N$/d$z$ measurements by treating DEEP2 as simply measuring the
velocity function at $z=1$ (the expected median redshift of the
survey).  This should provide the same constraints that would result
from likelihood maximization over redshift and velocity
simultaneously, but reduces the number of variables to be evaluated
and thereby speeds computation.  Thus, to determine the uncertainty
resulting from one of the three systematic effects we: 1) Choose a
``true'' value for the free parameter $\alpha$ of the systematic,
while using fiducial values for the other two.  This defines a single
probability distribution function (PDF) for galaxies (which for this
purpose we define as all objects with $v_c < 300$ km s$^{-1}$). 2)
Generate 500 Monte Carlo realizations of the velocity distribution of
DEEP2 galaxies by drawing an appropriate number of objects
(e.g. 10,000 for our ``best bet'' scenario) from this PDF.  3) For
each realization, we attempt to determine the input value of the free
parameter by likelihood maximization (using PDFs defined on a dense
grid in $\alpha$).  The distribution of the best-fit values from these
realizations defines the error in measuring $\alpha$; this may be
propagated into the uncertainty in the d$N$/d$z$ measurement.  We now
describe in detail the systematic effects we have investigated and the
models we have used for them:

\subsection{Baryonic Infall}

As the baryons within a dark matter halo cool and collapse to form a
galaxy, the mass distribution of the halo must respond.  This changes
the depth of the potential well, and thus the circular velocity in the
halo on galactic scales \cite{blumenthal86}.  Effectively, baryonic
infall remaps a set of dark matter halos to higher circular velocity.
Although the idea is not new, theoretical investigations of baryonic
infall are still improving \cite{bullock01}; we expect that they will
be significantly advanced by the completion of DEEP2.  We thus have
simply adopted a straightforward model from the literature (Gonzalez et
al. 2000).

That model defines a remapping of the velocity function
which combines an analytic treatment of baryonic infall with results
of N-body simulations.  There are three parameters in this
model: $\lambda$, the galaxy spin parameter; $c_{vir}$, the
concentration index for the Navarro, Frenk \& White dark matter halos
assumed by Gonzalez et al.; and $m_d$, the fraction of baryonic mass
that forms the disk.  The spin parameter $\lambda$ is well-determined
by N-body simulations; Gonzalez et al. assume $\lambda$=0.04, which
yields results essentially identical to integrating over the GIF
$\lambda$ distribution (Frenk et al. 2000).  The concentration index
$c_{vir}$ may also be determined from N-body simulations or
semi-analytically (Bullock et al. 2000); furthermore, the effects of
baryonic infall are essentially independent of $c_{vir}$ for $c_{vir}
\gtsim 5$, a range DEEP2 galaxies at $z\sim1$ should all fall
within. Based on N-body simulations, Gonzalez et al. give a relation
between $c_{vir}$ and $v_{circ}$ for $z=0$, $c_{vir}\approx 13 \times
\sqrt{v_{circ}/200}$~km s$^{-1}$; we take this as our guide,
with prefactor taken to be 8 rather than 13 (following the results of
Bullock et al. 2000).  This choice for our toy model is nearly
arbitrary; appropriate simulations would be used for baryonic infall
corrections to real data, but we do not require that for now.

However, the final component of the model, $m_d$, cannot be determined
so reliably from simulations or semi-analytic models.  We thus adopt
it as the free parameter in our baryonic infall model, or more
specifically, $m_0$, the prefactor in the relationship between $m_d$
and circular velocity (taken to be 0.1 by Gonzalez et al.).  This
parameter has a much stronger impact on the baryonic infall remapping
than $\lambda$ or $c_{vir}$, and will be least well-known from
simulations.  The impact of $m_0$ on baryonic infall-corrected velocity
functions is shown in Fig. \ref{bipdf}.

\subsection{Velocity Errors}

Observations of circular velocities of DEEP2 galaxies will, of course,
be subject to measurement error.  Furthermore, optical spectroscopy of
emmision-line objects will only determine the velocity of ionized gas
at some radius within a galaxy; it is unlikely that what is measured
will be identical to the dark halo circular velocity.  For instance,
Kobulnicky \& Gebhardt (1999) found that for galaxies at $z \sim 0$,
velocity measurements using O[II] and HI for the same galaxy have a
$\sim$20\% scatter.  Because the velocity function follows a very
steep power law ($n(v) dv \sim v^{-4}$) for galaxy-scale halos, such a
scatter can greatly change its shape and therefore potentially affect
d$N$/d$z$ measurements.  For the purposes of this paper we wish to
find if the amount of scatter may be determined from the observed
velocity function alone.

We have adopted a simple toy model for the error in velocity
measurements:\\ $\sigma_v=\sqrt{(25~\rm{km s^{-1}})^2+(f v_{circ})^2}$.
The first term arises from the instrumental resolution of DEIMOS,
which provides a fundamental limitation on the precision with which we
can measure small velocities.  The second term represents everything
else that causes scatter between observed velocities and $v_{circ}$
(after correction for baryonic infall).  The results of Kobulnicky et
al. suggest that the fractional error $f \sim 0.1$ locally (as the
20\% scatter they found includes the effects of resolution); $f \sim
0.2$ might be reasonable for galaxies at $z\sim1$.  The effect which
various values of $f$ have upon the velocity function is shown in Fig. \ref{velsigpdf}.

\subsection{Incompleteness}

Two types of incompleteness will affect attempts to use observable
galaxies to count dark matter halos halos: incompleteness due to
galaxies of a certain luminosity falling beyond the magnitude limit of
the survey, and incompleteness due to dark halos which do not contain
an observable galaxy.  The former (which we will hereafter refer to as
``luminosity incompleteness'') is easy to understand: given a
luminosity-linewidth relation and its scatter, one can directly
calculate the fraction of galaxies for a given linewidth that will
fall beyond the magnitude limit of the DEEP2 survey, and adjust the
velocity function accordingly.  On the other hand, knowing what
fraction of dark halos with a certain circular velocity host galaxies
above the surface brightness limit of a survey (or, for that matter,
contain a galaxy at all) requires a full model of galaxy formation.
It is not clear if this uncertainty, which we will refer to as ``halo
incompleteness'', is an issue for the $\sim L_{\star}$ galaxies DEEP2
will observe or not.  Unobserved low-surface brightness galaxies could
also be a problem for the $z \sim 0$ comparison samples required for a
d$N$/d$z$ measurement.  However, the local surveys now underway will
contain $5-20$ times as many galaxies as DEEP2, determining the impact
of incompleteness from the observed velocity function should if
anything be more effective than at $z\sim1$.

Lacking a model for halo incompleteness, we adopt luminosity
incompleteness as our toy model.  Although Bullock et al. (2001)
suggest that this effect will compromise measurement of d$N$/d$z$
using DEEP2 because the Tully-Fisher relation evolves with redshift,
that seems highly unlikely to be a problem.  With a sample of $\sim
10^4$ galaxies that have both magnitudes and linewidths, defining the
slope, zero point, scatter, and evolution of a luminosity-linewidth
relation should all be possible.  The effect of luminosity
incompleteness on the velocity function is then completely fixed, and
may be corrected for directly.  We demonstrate below that in fact we
could accurately determine luminosity incompleteness (in the absence
of halo incompleteness) using the velocity function alone.  However,
since we lack sufficient simulations, we presume that the impact of
halo incompleteness on the velocity function would be generally
similar, so that we can assess the impact of halo incompleteness by
evaluating how well we can determine luminosity incompleteness from
the velocity function alone.  This should be generically true; it is
likely that all high-circular velocity halos will contain a galaxy,
and one above any surface brightness limitations of the survey at
that, while smaller halos may well not.  Thus, the signature of the
two sorts of incompleteness is similar: full completeness for
high-velocities halos, rolling over to incompleteness at lower
velocities.  The issues involved in halo incompleteness should be much
better understood by the time the DEEP2 survey is completed, as
simulations and semianalytic techniques should only improve.

For our toy model of incompleteness, we follow the results of Vogt et
al. (1997) that the slope and scatter of the rest-frame $B$
Tully-Fisher relation appear to remain unchanged to $z\sim1$.  We
take as a free parameter the zero point of the relation, or
equivalently $v_{50}$, the circular velocity (after baryonic infall)
at which observations to the DEEP2 magnitude limit ($I_{AB}\sim 23.5$)
will be 50\% complete at $z=1$.  We thus obtain a Tully-Fisher
relation:
\begin{equation}
M_B = -7.48 \log_{10}{v_c} +40.23  - C,
\end{equation}
where $C$ combines the $K$ correction (small for conversion of $B$ at
z$\sim$0 to $I$ at $z\sim 1$), the Hubble constant (as $-5
\log_{10}{h}$) and evolution of the Tully-Fisher zero point; the results
of Vogt et al. suggest $C\sim 0.2\pm 0.3$ for an $h=0.75$ LCDM model,
corresponding to $v_{50}\sim 160 \pm 15$ km s$^{-1}$.  They found
that the 0.65 mag dispersion in the Tully-Fisher relation at high
redshift is consistent with being due to a 0.4 mag intrinsic scatter,
a 0.2 mag scatter due to magnitude measurement errors, and a 0.47 mag
scatter due to velocity errors.  We apply our model incompleteness to
the velocity function before adding noise, so only the first two
elements of the scatter are relevant.  The effect of varying $v_{50}$ on
the velocity function is shown in Fig. \ref{incpdf}.

\subsection{Resulting Errors}

Given these toy models, we have generated Monte Carlo realizations of
the resulting velocity functions as the parameters are varied, as
described above.  The net effect of these systematics in a standard
scenario ($m_0$=0.1, $f=0.2$, $v_{50}=160$ km s$^{-1}$) is shown in
Fig. \ref{fullpdf}.  We expect that one-fourth to one-half of the
50,000 galaxies for which DEEP2 should obtain redshifts will provide
velocity measurements.  We therefore have used from 5,000 galaxies
(corresponding to the total number of objects in the deeper 3HS) to
20,000 galaxies in our Monte Carlo realizations of the velocity
function.  Based upon the simulations, we have compiled error budgets
for a determination of the volume element in three scenarios: a
pessimistic one in which there are only 5,000 objects with linewidth
information and the free parameters of the systematic effects take
values which result in the greatest errors; a ``best bet'' scenario in
which there are 10,000 objects and the free parameters have
intermediate values; and an optimistic scenario in which there are
20,000 objects and the free parameters take favorable values.  We
present the resulting error budgets for a measurement of apparent
density within the entire DEEP2 survey volume in Table 1.\footnote{We
have not considered the degree to which systematics may be measured or
eliminated by study of the local velocity function, making our error
budget a conservative one.  Baryonic infall models, in particular,
could be further constrained by $z\sim0$ observations, reducing our
dominant systematic uncertainty.}  In every case, the distributions of
the residual errors after correction for systematics showed no major
non-Gaussianities.

\section{Cosmological Constraints}

The DEEP2 Redshift survey will not simply measure d$N$/d$z$ at one
redshift, but instead over the entire range $0.7<z<1.5$.  To determine
the resulting constraints upon cosmological parameters, we must take
this into account explicitly.  Because both cosmic variance and the
residual errors from systematic effects after correction are
essentially Gaussian, we may use a $\chi^2$ paradigm.  To simplify, we
presume that DEEP2 measures d$N$/d$z$ in eight redshift bins, each
covering 0.1 in $z$.  The observed abundance in each bin will be
affected by Poisson variance, by the residual errors from systematic
effects after correction, and by cosmic variance.

To determine the Poisson variance in each bin, we must know the number
of galaxies within it.  To estimate this, we use a fit to the
photometric redshift distribution of galaxies in the Hubble Deep
Fields to the 1HS magnitude limit \cite{gwynhartwick,lanzetta}, which
yields d$N$/d$z \propto (1+z)^{-3.25}$, and the total number of
galaxies appropriate to each error scenario.  To first order, residual
systematic errors will affect all redshift bins in the same way.
Thus, for these errors we use the same fractional error
in d$N$/d$z$ for all redshift bins and presume it is completely
correlated amongst them.  Finally, given a cosmological model, we may
calculate the cosmic variance within each redshift bin as in section
2; we treat it as completely independent between bins.  In actuality,
the number of objects in adjoining bins will be slightly correlated, but
this is negligible compared to the uncorrelated part of the variance
(or to the residual errors from systematics; the correlated part of
the cosmic variance between our redshift bins is less than 10\% of the
total).

Given these definitions, the covariance matrix for our redshift bins
is completely determined; we list the values used (for an LCDM model)
in Table 2.  We may then calculate $\chi^2$ between any model and some
nominal, ``true'' model (e.g. LCDM: $\Omega_m=0.3$, $\Omega_Q=0.7$,
$w=-1$).\footnote{We use the extension of $\chi^2$ to a multivariate
distribution with covariance: $\chi^2=(\bf{n-n_0})^T V^{-1}
(\bf{n-n_0})$, where $\bf{n}$ is the vector of observations,
$\bf{n_0}$ is the vector of true values, and $\bf{V}$ is the
covariance matrix for $\bf{n_0}$.  The elements of $\bf{V}$ are listed
in Table 2.}  Observed results should be distributed as $\chi^2$ with
two degrees of freedom, so the results can be directly translated into
statistical confidence contours.  In
Figs. \ref{mlcontour}--\ref{mwcontour7} we show the results for all
three error budgets.  Fig. \ref{syscv} shows the separate effects of
cosmic variance and residual systematics.  For the optimistic, best
bet, and pessimistic scenarios, the error in a measurement of $w=-0.7$
will be 0.069, 0.086, or 0.11 if $\Omega_m$ is known to $\pm 0.025$,
as compared to 0.051 if Poisson errors dominated.  The worse the
errors on $\Omega_m$, the less relative degradation cosmic variance
and residual systematics will cause: if the error in $\Omega_m$ were
0.01, the resulting variance would be $2.2-6.5\times$ as large as
predicted from Poisson statistics, while if it were 0.05, precision
would be only $15-60\%$ worse than the Poisson prediction.

In Fig. \ref{bestcontour}, we present the best bet contours along with
predicted or recent results for a variety of cosmological tests.
Comparison to Fig. 3 of ND00 shows that even an optimistic but
realistic error budget would yield poorer constraints than we
previously predicted based on purely statistical errors.  On the other
hand, the measurement should yield much stronger limits on the dark
energy than any other method available on the same timescale, at
minimal marginal cost as it is a byproduct of a ground-based redshift
survey that has been designed to address a number of issues in galaxy
evolution and cosmology.  If SNAP results are dominated by statistical
uncertainties, they would yield cosmological constraints a factor of
$2-3\times$ stronger than a d$N$/d$z$ measurement based on the DEEP2
galaxy velocity function alone.  In the long term, the DEEP2 results
would remain a valuable check on SNAP and other methods, however, as
they will be subject to very different systematic errors than other
cosmological tests.  DEEP2 will also provide other complementary
measurements of cosmological parameters through a variety of methods
(e.g. by studying the abundance and velocity function of galaxy
clusters within the survey volume; see Newman et al. 2001).

In summary, we have shown that the uncertainties of Newman \& Davis
(2000) were overly optimistic by a factor of $\sim 2$, as we ignored
the effects of cosmic variance and had not yet tested the assumption
that systematic effects could be found and corrected for based on the
observed velocity function.  We have demonstrated that this procedure
can be implemented, though the residual errors from the correction
procedure somewhat exceed the Poisson errors in a d$N$/d$z$
measurement.  However, we have found no evidence that the test is
compromised either by cosmic variance or any of the systematic effects
considered.  We must note also that our analysis has been
conservative; for instance, we have not assumed that observations of
the velocity function at $z\sim 0$ provides any constraints on
systematic errors, nor have we considered the fact that comparison to
larger-area multicolor imaging or redshift surveys could allow us to
normalize out cosmic variance in the DEEP2 fields.  With sufficient
attention to detail, the classical d$N$/d$z$ test can provide much
stronger constraints on dark energy than have ever been available
before.

\acknowledgements

We wish to thank Michael Turner, whose challenge to produce a
realistic error budget inspired this work, and Richard Ellis, whose
inquiry about the importance of cosmic variance caused us to
investigate the issue.  Andrew Jaffe kindly provided the
BOOMERANG/MAXIMA likelihood contour plotted here.  We also thank
Alison Coil, Christian Marinoni, and Martin White for helpful
discussions.  This material is based upon work supported by the
National Science Foundation under Grant No. AST-0071048.  This work
was also made possible by equipment donated by Sun Microsystems.




\clearpage
\begin{figure*}
\epsscale{2}\plotone{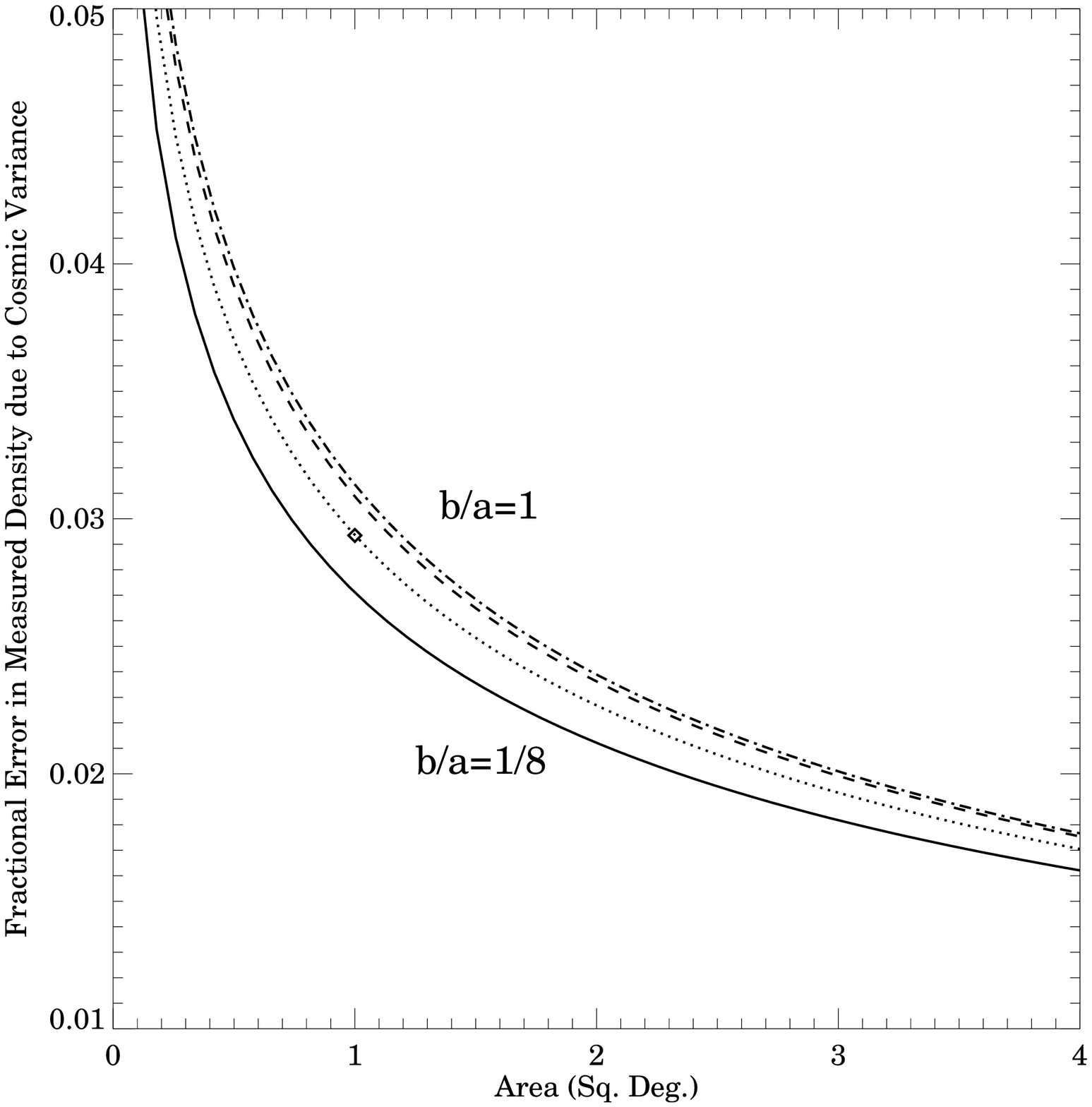}
\caption{The fractional cosmic variance in a density measurement as a
function of the area on the sky surveyed per field (4 fields assumed)
in a volume extending from $0.7<z<1.5$.  Curves are plotted for axis
ratios $b/a=1/8,1/4,1/2$ and 1.  The planned geometry for the
DEEP2 survey is indicated with a diamond.\label{areaplot} }
\end{figure*}

\clearpage
\begin{figure*}
\epsscale{2}\plotone{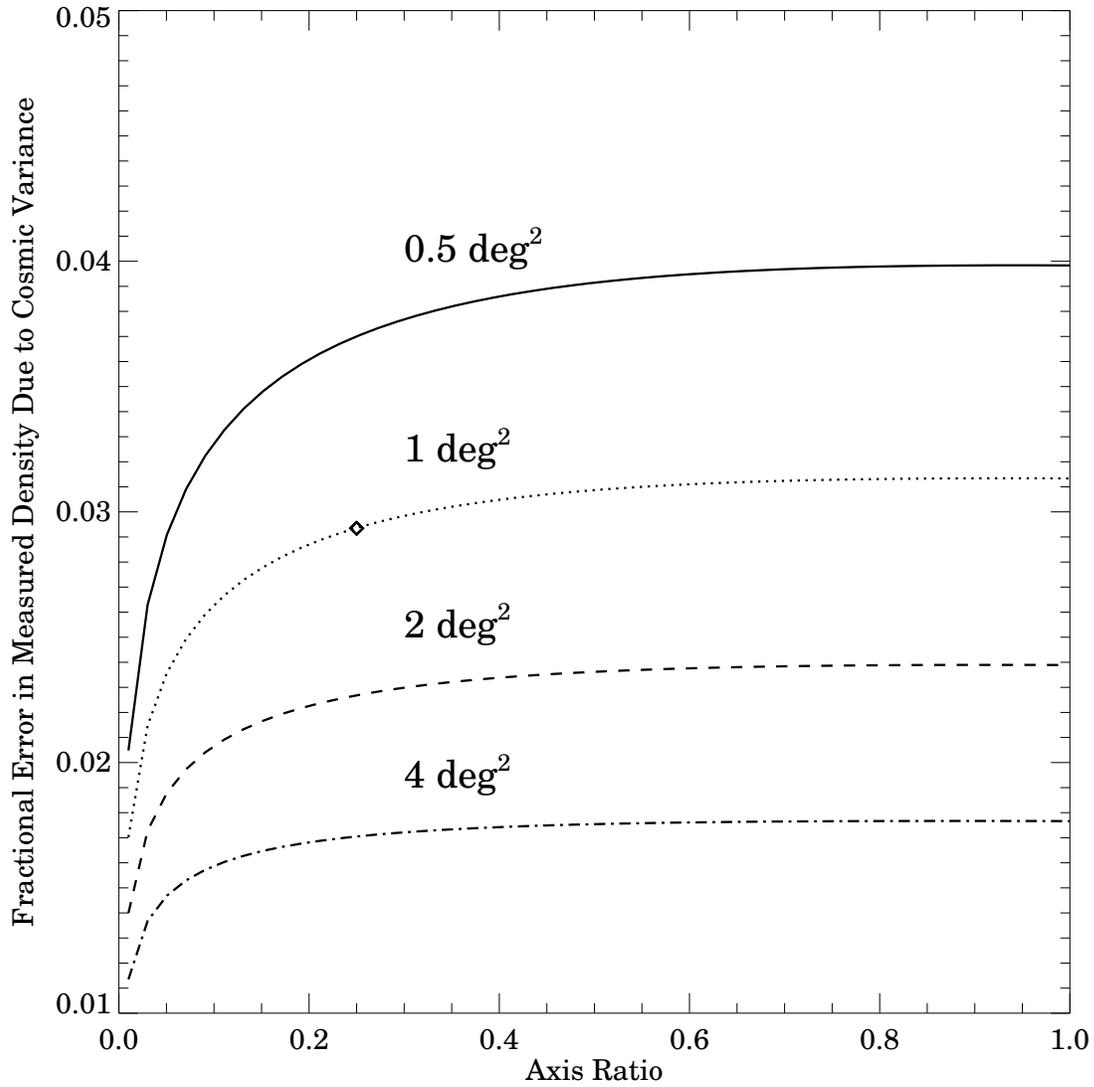}
\caption{As Fig. \ref{areaplot}, but as a function of axis ratio on the sky
rather than area.  Save in the most elongated cases, the amount of area surveyed per field is much more important than the axis ratio. \label{axisplot} }
\end{figure*}

\clearpage
\begin{figure*}
\epsscale{2}\plotone{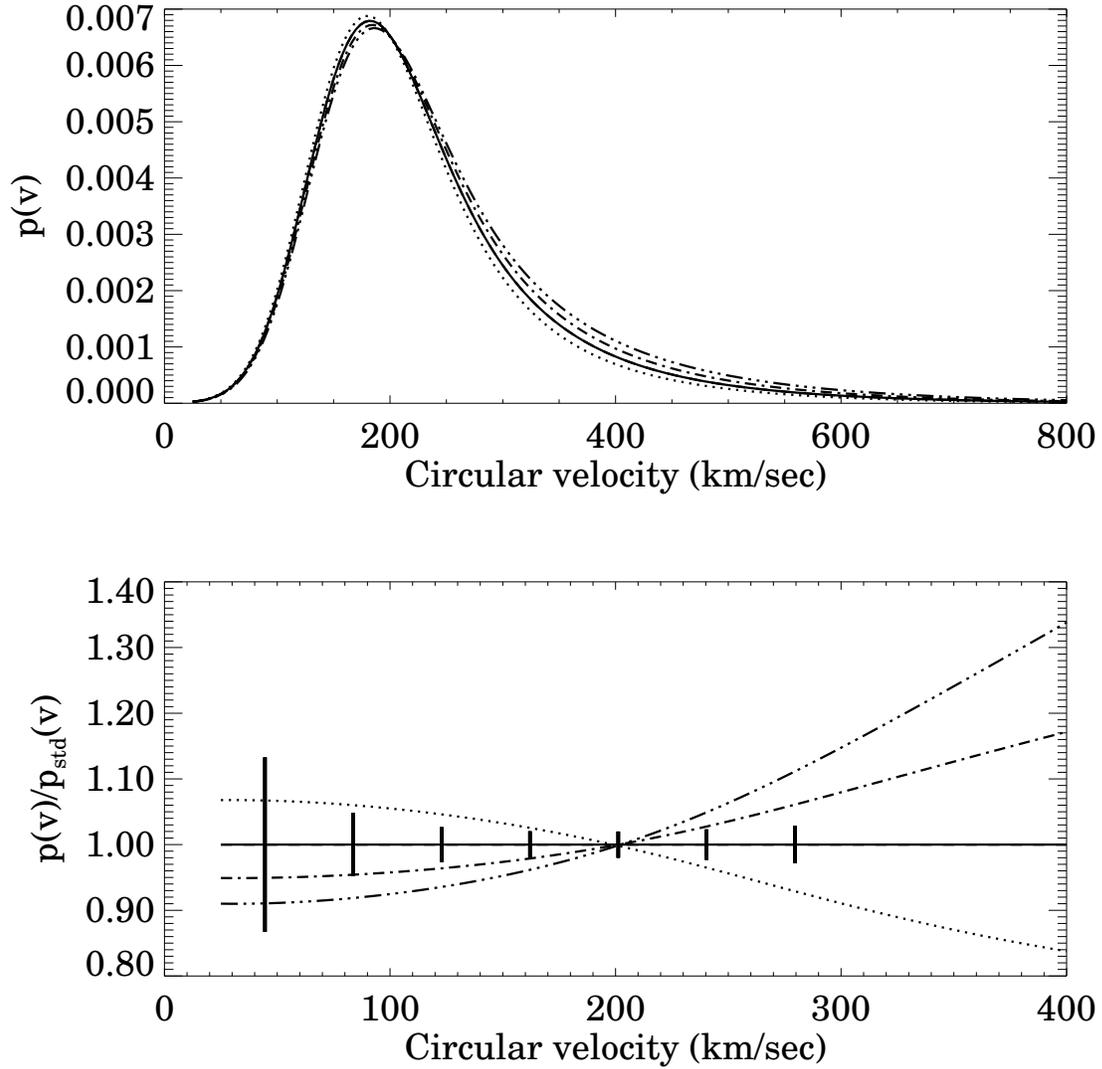}
\caption{(top) The PDF for the circular velocity of dark matter halos
at $z=1$ as the disk fraction parameter $m_0$ is varied from 0.05
(dotted) to 0.2 (dot-dot-dot-dash).  (bottom) The ratio of the PDF for
varying values of $m_0$ to that in the fiducial model.  Also plotted
are expected Poisson error bars for that model if the velocity function is
divided into seven bins. Compare to Figs. \ref{velsigpdf} and
\ref{incpdf}; the signatures of the three systematic effects differ
strongly.\label{bipdf} }
\end{figure*}

\clearpage
\begin{figure*}
\epsscale{2}\plotone{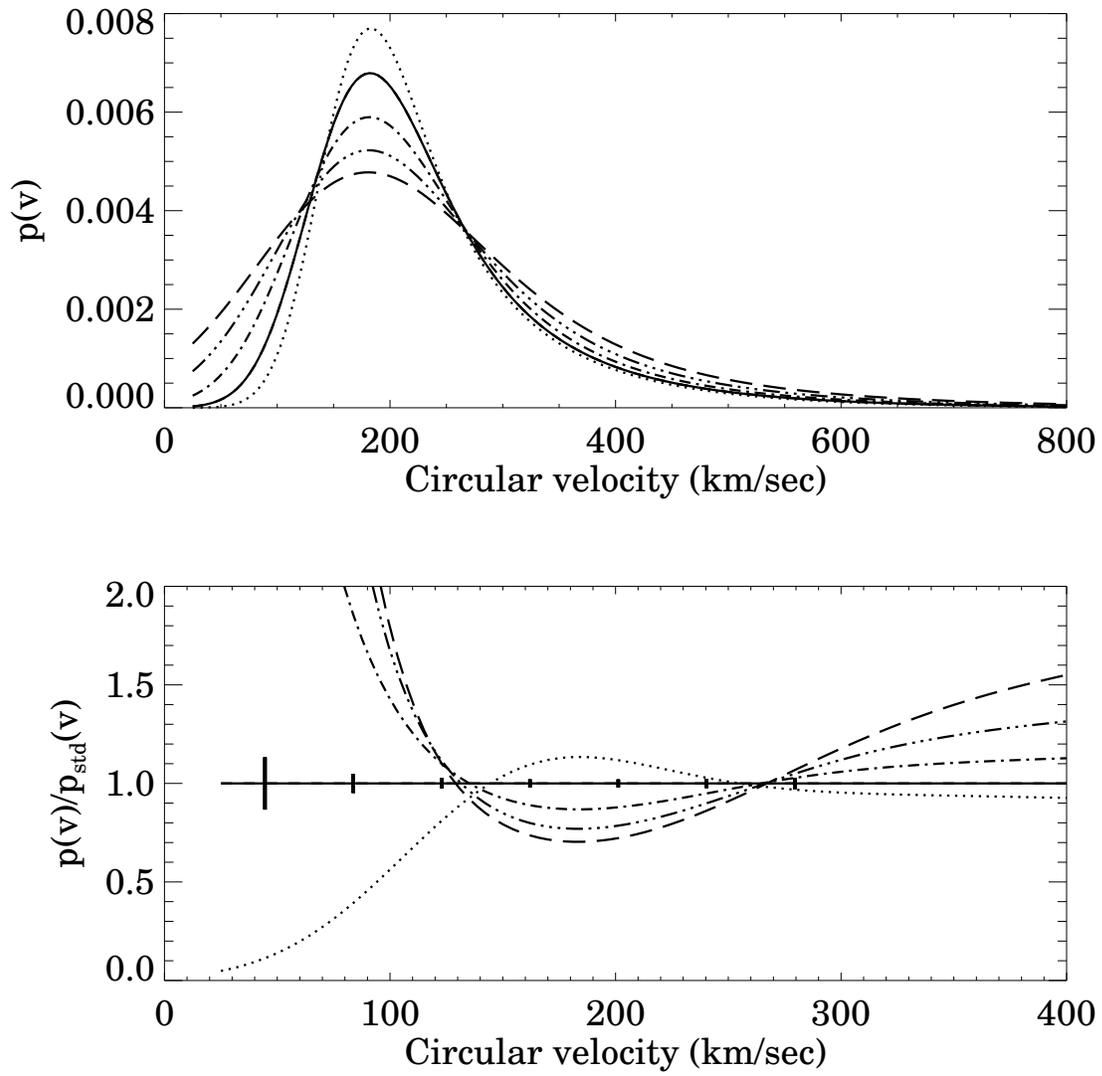}
\caption{(top) As Fig. \ref{bipdf}, but now the velocity error
parameter $f$ is varied from 0.1 (dotted) to 0.4 (dot-dot-dot-dash).
(bottom) The ratio of the PDF for varying values of $f$ to that in the
fiducial model. The signature is much larger than the expected
errors.\label{velsigpdf} }
\end{figure*}

\clearpage
\begin{figure*}
\epsscale{2}\plotone{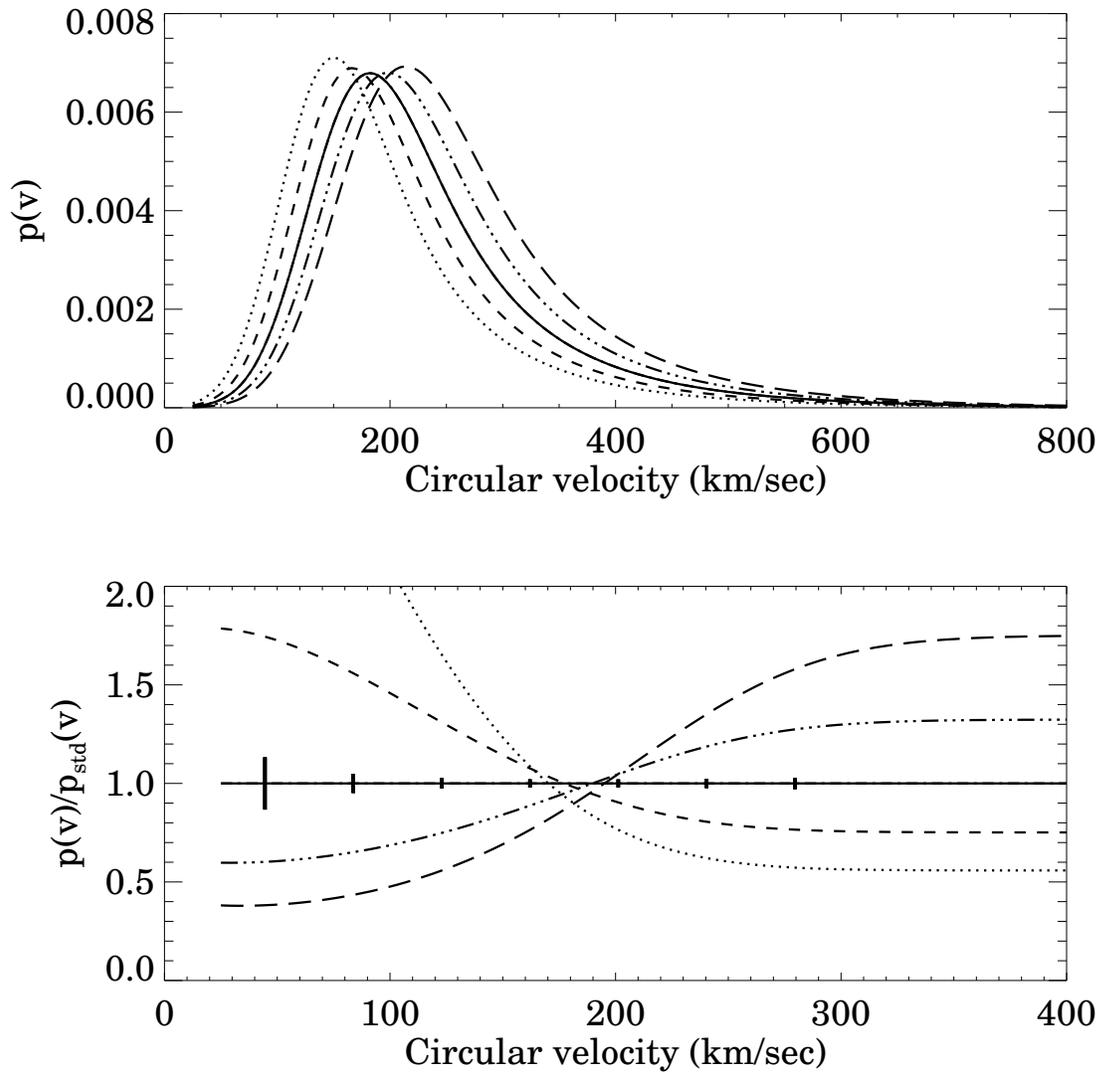}
\caption{(top) As Fig. \ref{bipdf}, but now the 50\% incompleteness velocity $v_{50}$ is varied from 130 (dotted) to 190 (dot-dot-dot-dash) km s$^{-1}$.
(bottom) The ratio of the PDF for varying values of $v_{50}$ to that
in the standard model. Again, the signature is quite strong.\label{incpdf} }
\end{figure*}

\clearpage
\begin{figure*}
\epsscale{2}\plotone{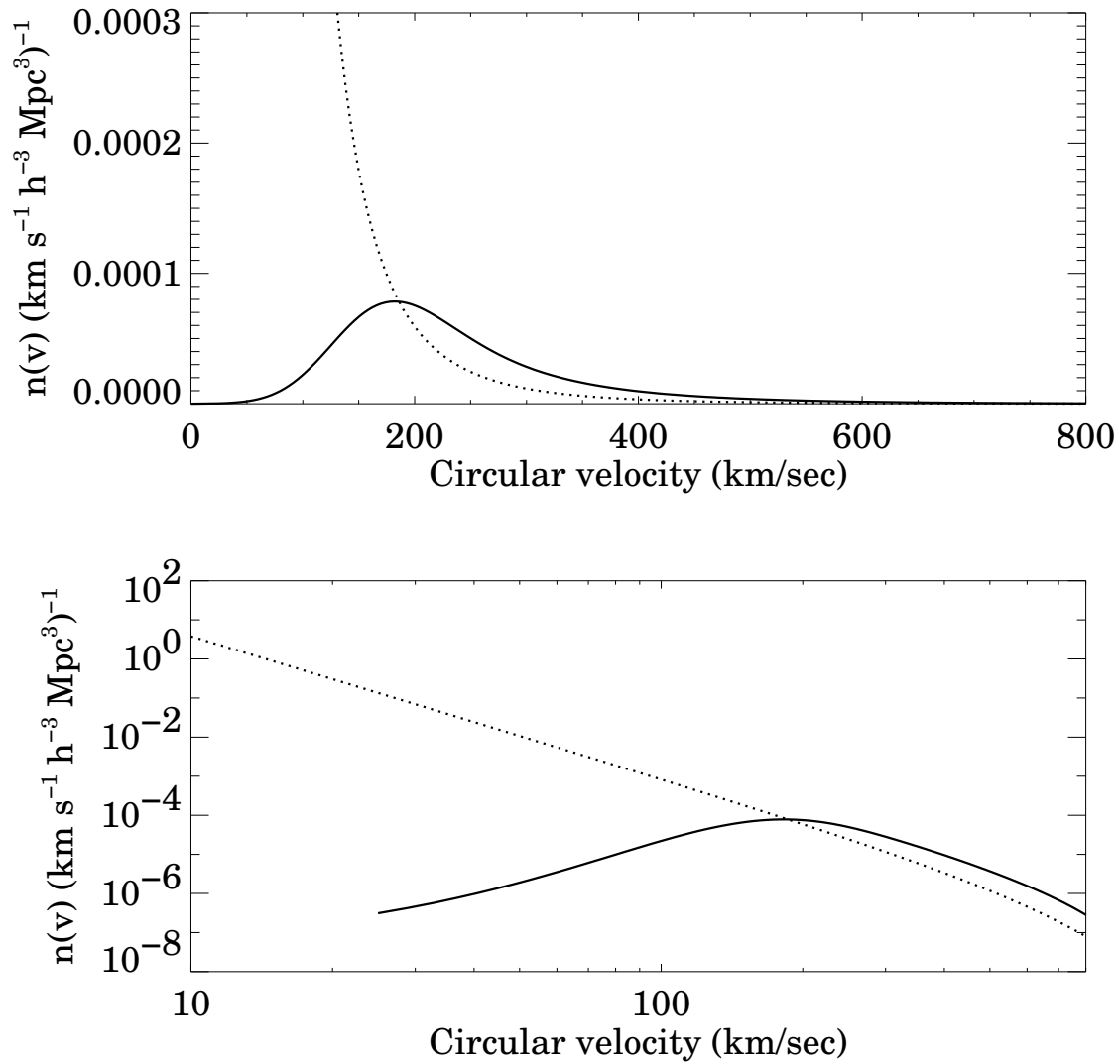}
\caption{The differential circular velocity function for dark matter halos at $z=1$, before (dotted) and after (solid) the systematic effects described in $\S$ 3 are applied.  For this plot, the standard values of the free parameters ($m_0=0.1$, $f=0.2$, $v_{50}=160$ km s$^{-1}$) were used.  \label{fullpdf} }
\end{figure*}
\clearpage
\begin{figure*}
\epsscale{2}\plotone{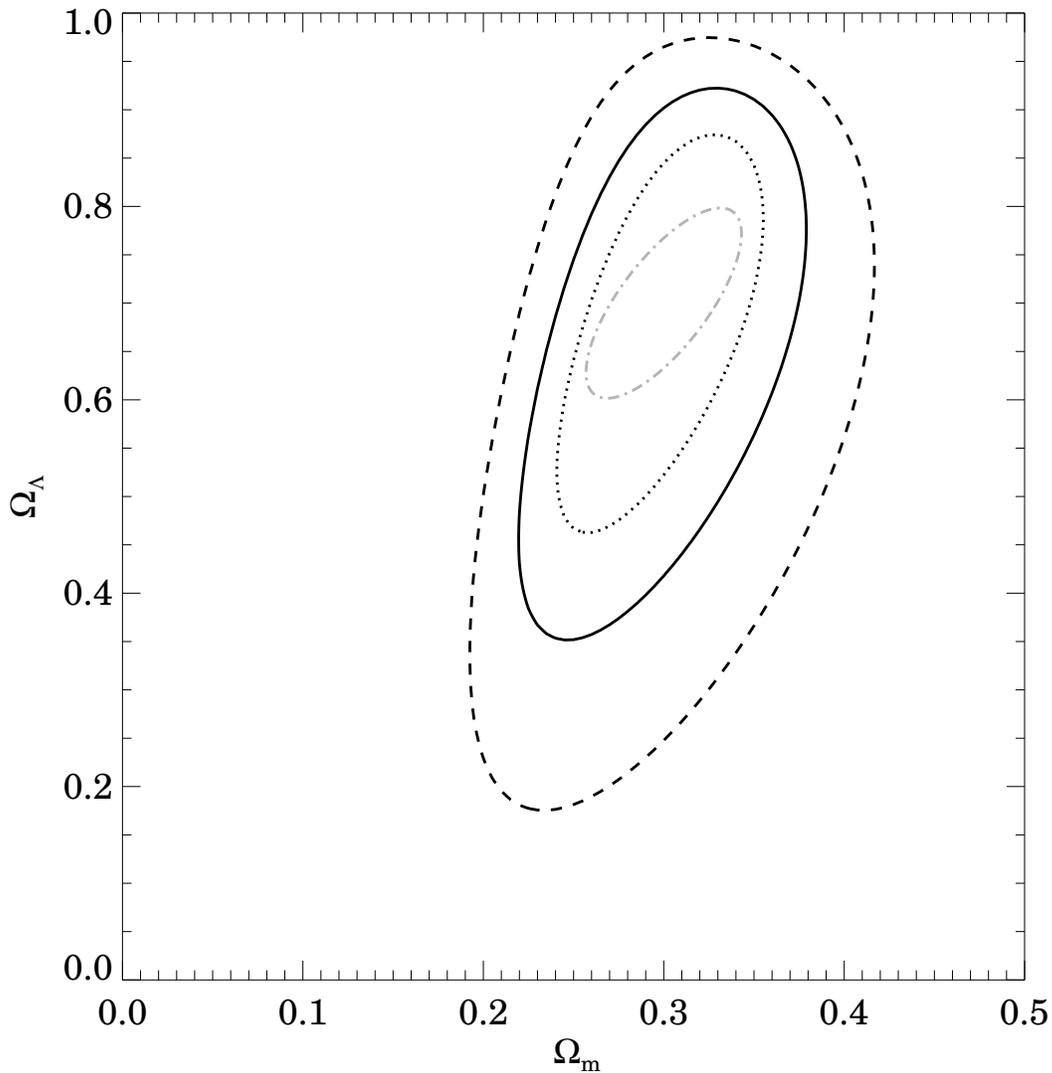}
\caption{Black curves: 95\% confidence constraints in the
$\Omega_m-\Omega_{\Lambda}$ plane resulting from our three error
scenarios in an $\Omega_m=0.3$, $\Omega_{\Lambda}=0.7$ model:
optimistic (dotted), ``best bet'' (solid), and pessimistic (dashed).
Also plotted for comparison is the target 95\% statistical uncertainty
for the SNAP project (grey, dot-dashed curve).  \label{mlcontour} }
\end{figure*}

\clearpage
\begin{figure*}
\epsscale{2}\plotone{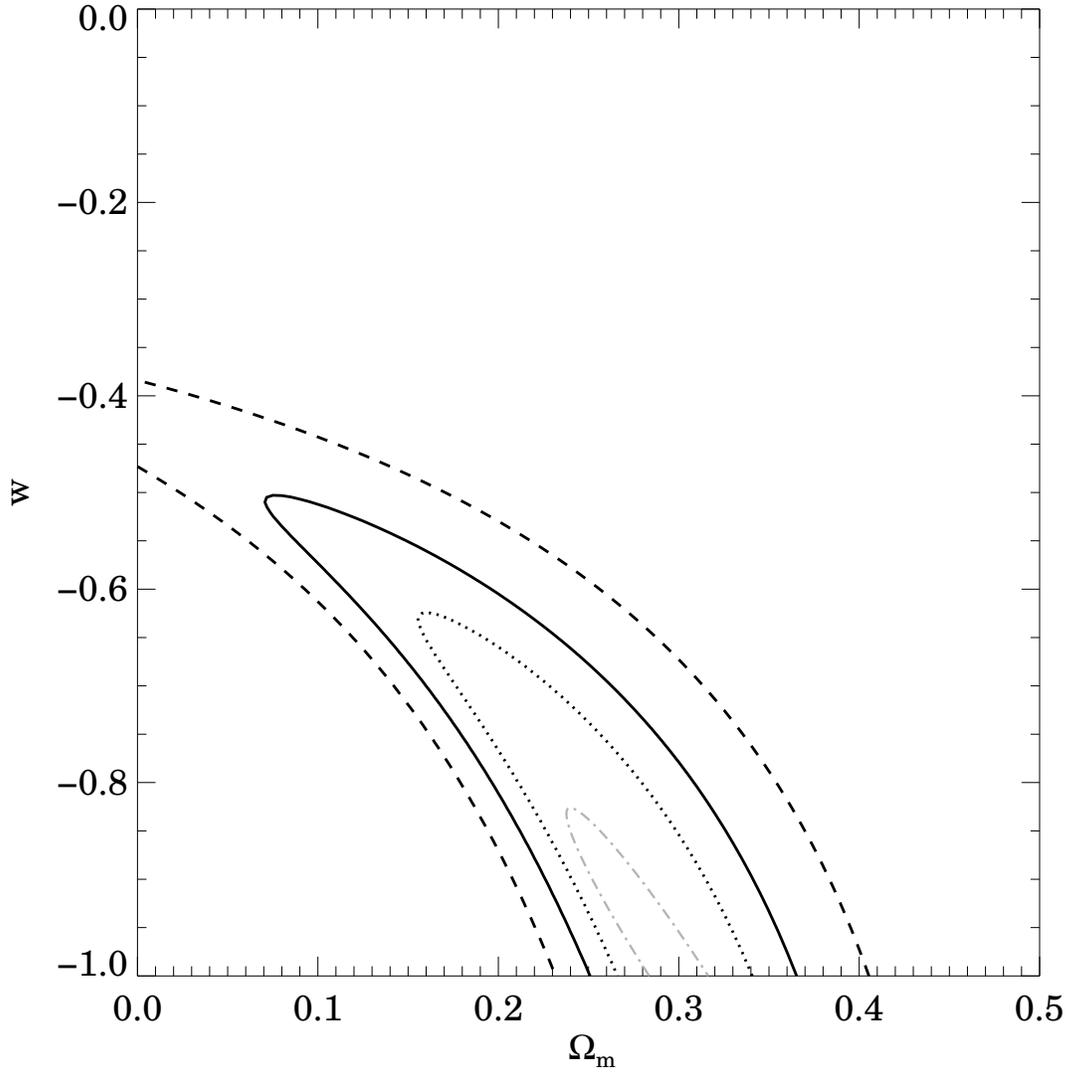}
\caption{Black curves: 95\% confidence constraints in the $\Omega_m-w$ plane resulting from our three error scenarios in an $\Omega_m=0.3$, $\Omega_{Q}=0.7, w=-1$ model: optimistic (dotted), ``best bet'' (solid), and pessimistic (dashed).  Also plotted is the target 95\% statistical uncertainty for the SNAP project for comparison (grey, dot-dashed curve).  \label{mwcontour} }
\end{figure*}

\clearpage
\begin{figure*}
\epsscale{2}\plotone{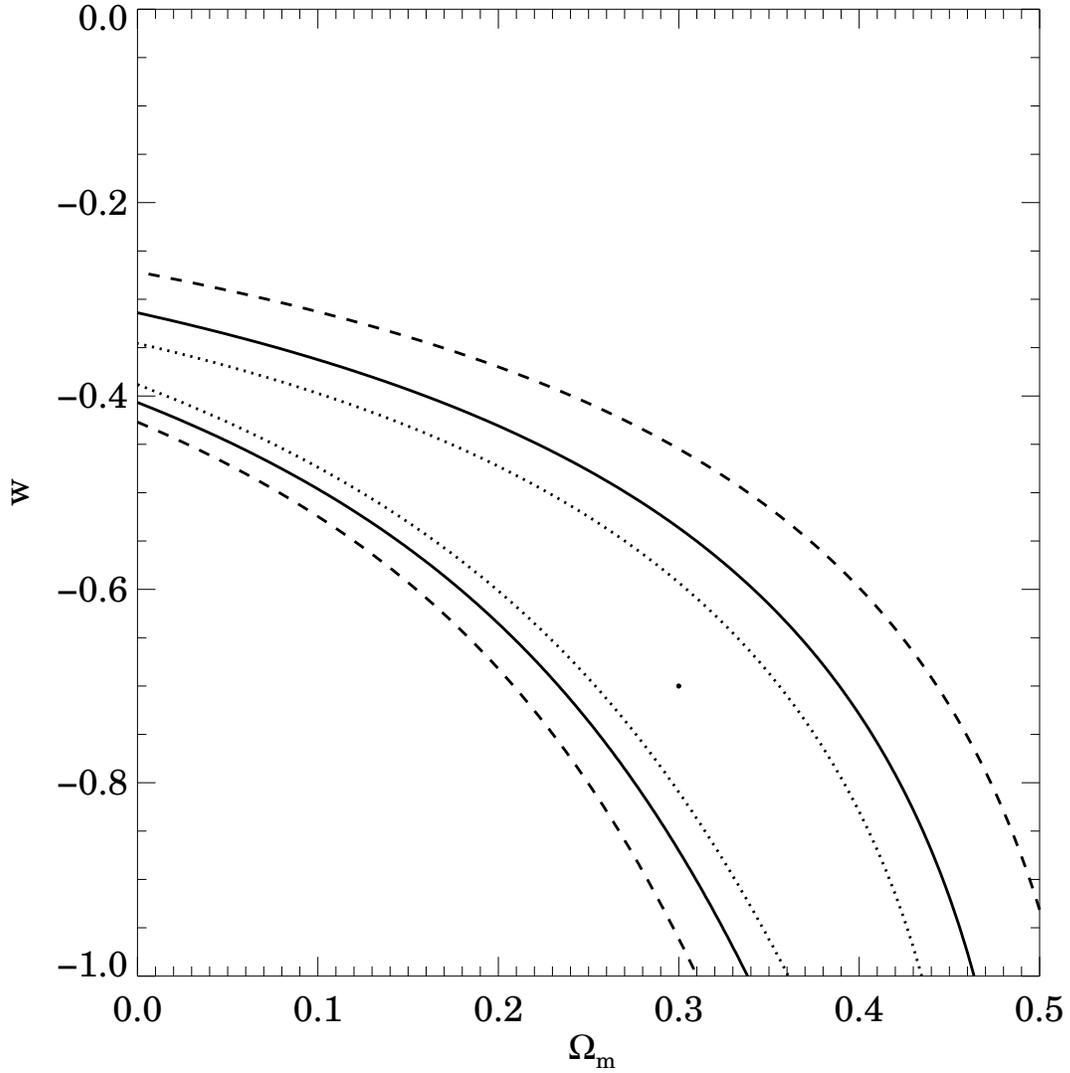}
\caption{95\% confidence constraints in the $\Omega_m-w$ plane resulting from our three error scenarios n an $\Omega_m=0.3$, $\Omega_{Q}=0.7, w=-.7$ model: optimistic (dotted), ``best bet'' (solid), and pessimistic (dashed).    \label{mwcontour7} }
\end{figure*}

\clearpage
\begin{figure*}
\begin{footnotesize}
\renewcommand{\baselinestretch}{1.0}
\epsscale{0.9}
\epsscale{2}\plotone{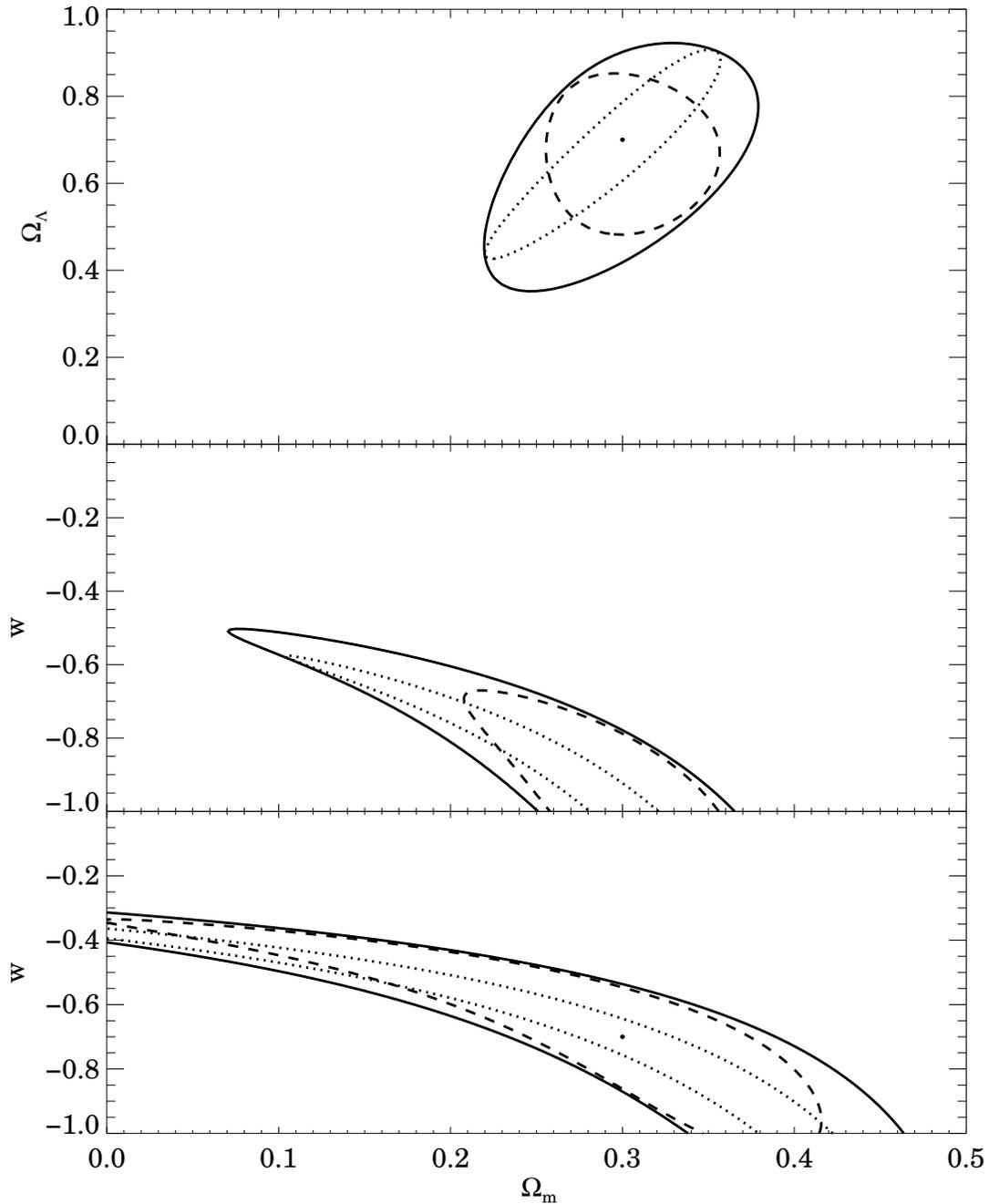}
\caption{{ The best bet contours from the previous three figures
(solid black curves), along with curves indicating what the
constraints with zero systematic errors (dotted curves) or zero cosmic
variance (dashed curves).  (Top) Constraints in the
$\Omega_m-\Omega_{\Lambda}$ plane for an $\Omega_m=0.3$,
$\Omega_{\Lambda}=0.7$ model.  (Middle) Constraints in the
$\Omega_m-w$ plane for a model with $\Omega_m=0.3$, $w=-1$.  (Bottom)
As the middle panel, but for a model with $w=-0.7$.  In most cases,
the effects of both cosmic variance and systematic effects are of
comparable importance.}
\label{syscv} }
\end{footnotesize}
\end{figure*}

\clearpage
\begin{figure*}
\begin{footnotesize}
\renewcommand{\baselinestretch}{1.0}
\epsscale{0.8}
\epsscale{2}\plotone{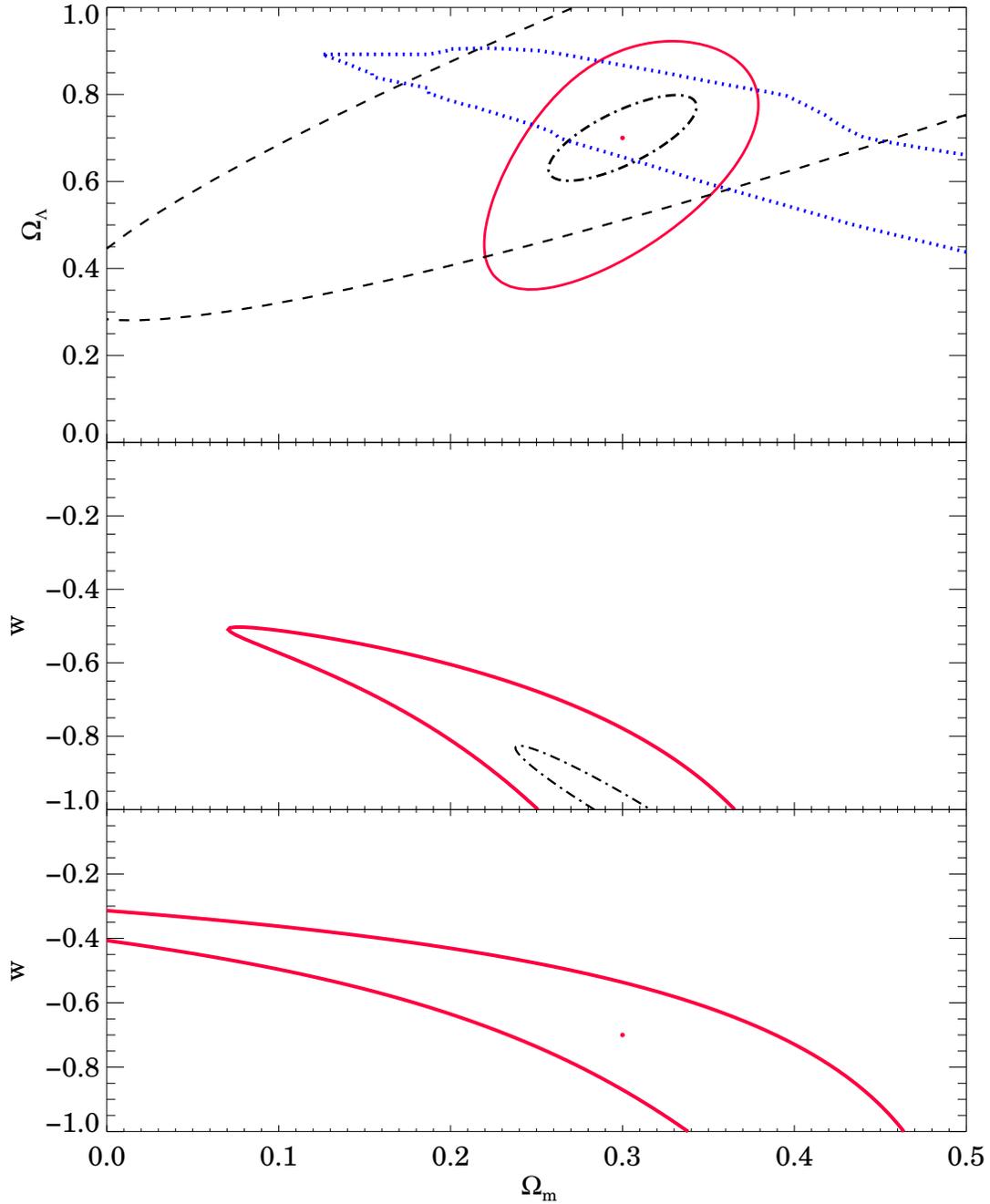}
\caption{{ The best bet contours from the previous three figures
(solid red curves), along with other current or near-future
measurements.  (Top) Plotted constraints in the
$\Omega_m-\Omega_{\Lambda}$ plane include the current 68\% SNe Ia
results (Perlmutter et al. 1999; large black dashed curve) and BOOMERANG/MAP measurements
(Jaffe et al. 2000; blue dotted curve), along with (for an $\Omega_m=0.3$,
$\Omega_{\Lambda}=0.7$ model) the 95\% SNAP target statistical
uncertainty (black dot-dash curve).  
(Middle) Potential measurements in the $\Omega_m-w$ plane
for a model with $\Omega_m=0.3$, $w=-1$.  In addition to our ``best
bet'' scenario, the SNAP target statistical uncertainty is shown (black dot-dashed curve).
(Bottom) As the middle panel, but for a model with $w=-0.7$.}
\label{bestcontour} }
\end{footnotesize}
\end{figure*}

\clearpage
\begin{deluxetable}{l|cl|cl|cl}
\tablecolumns{7}
\tablewidth{0pc}
\footnotesize
\tablenum{1}
\tablecaption{Error Budgets for Density Measurements from DEEP2}
\tablehead{
\multicolumn{1}{c}{} & \multicolumn{6}{c}{Error Scenario:} \\
 \multicolumn{1}{c}{} & \multicolumn{2}{c}{Pessimistic} & \multicolumn{2}{c}{Best Bet} & \multicolumn{2}{c}{Optimistic} \\
\multicolumn{1}{c}{Error Source} & \multicolumn{6}{c}{Fractional Error (Parameter Value)} 
}
\startdata
Counting Statistics & 0.014 &(N=5000) & 0.010 &(N=10000) & 0.007 &(N=20000)\\
Cosmic Variance & 0.022 &($\sigma_8=0.723$)& 0.018 &($\sigma_8=0.600$)& 0.014 &($\sigma_8=0.474$)\\
Baryonic Infall\tablenotemark{1} & 0.080 &($m_0=0.05$) & 0.050 &($m_0=0.1$)& 0.033 &($m_0=0.2$)\\
Velocity Errors\tablenotemark{1} & 0.012 &($f=0.4$)& 0.002 &($f=0.2$) & 0.001 &($f=0.2)$\\
Incompleteness\tablenotemark{1} & 0.021 &($v_{50}=175$)& 0.014 &($v_{50}=160$)& 0.009 &($v_{50}=145$)\\
\enddata
\tablenotetext{1}{Residual error when the observed velocity function is used to measure and remove the effect}
\end{deluxetable}

\clearpage
\begin{deluxetable}{cccc}
\tablecolumns{4}
\tablewidth{0pc}
\footnotesize
\tablenum{2}

\tablecaption{Covariance Matrix Elements \label{table2}}
\tablehead{
 \multicolumn{1}{c}{Element\tablenotemark{1}} & \multicolumn{1}{c}{Pessimistic} & \multicolumn{1}{c}{Best Bet} & \multicolumn{1}{c}{Optimistic} \\
}
\startdata
1,1 & 0.00313 $n_1^2$ & 0.00620 $n_1^2$ & 0.01203 $n_1^2$ \\
2,2 & 0.00307 $n_2^2$ & 0.00611 $n_2^2$ & 0.01196 $n_2^2$ \\
3,3 & 0.00303 $n_3^2$ & 0.00608 $n_3^2$ & 0.01197 $n_3^2$ \\
4,4 & 0.00303 $n_4^2$ & 0.00609 $n_4^2$ & 0.01205 $n_4^2$ \\
5,5 & 0.00304 $n_5^2$ & 0.00614 $n_5^2$ & 0.01220 $n_5^2$ \\
6,6 & 0.00308 $n_6^2$ & 0.00623 $n_6^2$ & 0.01241 $n_6^2$ \\
7,7 & 0.00313 $n_7^2$ & 0.00635 $n_7^2$ & 0.01267 $n_7^2$ \\
8,8 & 0.00320 $n_8^2$ & 0.00649 $n_8^2$ & 0.01298 $n_8^2$ \\
$i,j,i\neq j$ & 0.00123 $n_i n_j$ & 0.00305 $n_i n_j$ & 0.00719 $n_i n_j$ \\
\enddata
\tablenotetext{1}{These are elements of the covariance matrix for $n_i$,
where $n_i$ is the number of objects in the $i$th redshift bin.  An
LCDM model with $\Omega_m=0.3, \Omega_{\Lambda}=0.7, w=-1$ has been
used.  The first bin extends from $z=0.7$ to 0.8, the second from 0.8
to 0.9, etc.  If $f_{CV}$ is the fractional error from cosmic variance
in a bin, $f_{count}$ that from counting statistics, and $f_{sys}$
that from residual systematics (combined in quadrature), the diagonal
elements of the covariance matrix will be
$(f_{CV}^2+f_{count}^2+f_{sys}^2)n_i^2$, while the off-diagonal
elements will be $f_{sys}^2 n_i n_j$.}

\end{deluxetable}
\end{document}